\documentclass[conference]{IEEEtran}
\IEEEoverridecommandlockouts
\usepackage{cite}
\usepackage{amsmath,amssymb,amsfonts}
\usepackage{algorithmic}
\usepackage{graphicx}
\usepackage{textcomp}
\usepackage{xcolor}
\def\BibTeX{{\rm B\kern-.05em{\sc i\kern-.025em b}\kern-.08em
    T\kern-.1667em\lower.7ex\hbox{E}\kern-.125emX}}

\usepackage {diagbox}
\usepackage{ulem} 
\def\cG{{\mathcal G}}
\def\cL{{\mathcal L}}

\def\cT{{\mathcal T}}
\def\F{{\mathbb F}}

\newcommand{\vect}[1]{\mbox{\boldmath $#1$}}

\begin{document}

\title{
    Efficient pooling designs and screening performance in group testing for two type defectives\\

    \thanks{
        This work was supported by JSPS KAKENHI Grant Number JP22K11943.}
}

\author{
    \IEEEauthorblockN{1\textsuperscript{st} Hiroyasu Matsushima}
    \IEEEauthorblockA{
        \textit{Data Science and AI Innovation} \\
        \textit{Reseach Promotion Center}\\
        \textit{Shiga University}, Hikone, Japan\\
        0000-0001-7301-1956
    }
    \and
    \IEEEauthorblockN{2\textsuperscript{nd} Yusuke Tajima}
    \IEEEauthorblockA{
        \textit{Data Science and AI Innovation} \\
        \textit{Reseach Promotion Center}\\
        \textit{Shiga University}, Hikone, Japan\\
        yusuke-tajima@biwako.shiga-u.ac.jp
    }
    \and 
    \IEEEauthorblockN{3\textsuperscript{rd} Xiao-Nan Lu}
    \IEEEauthorblockA{
        \textit{Department of Electrical, Electronic}\\
        \textit{and Computer Engineering}\\
        \textit{Gifu University}, Gifu, Japan \\
        0000-0001-7881-8505
    }    
    \and
    \IEEEauthorblockN{4\textsuperscript{th} Masakazu Jimbo}
    \IEEEauthorblockA{
        \textit{Center for Training Professors in Statistics} \\
        \textit{The Institute of Statistical Mathematics}\\
        Tokyo, Japan \\
        jimbo@ism.ac.jp
    }
}

\maketitle

\begin{abstract}
Group testing is utilized in the case when we want to find a few defectives among large amount of items. 
Testing $n$ items one by one requires $n$ tests, but if the ratio of defectives is small, group testing is an efficient way to reduce the number of tests. 
Many research have been developed for group testing for a single type of defectives. 
In this paper, we consider the case where two types of defective A and B exist. 
For two types of defectives, we develop a belief propagation algorithm to compute marginal posterior probability of defectives. 
Furthermore, we construct several kinds of collections of pools in order to  test for A and B. 
And by utilizing our belief propagation algorithm, we evaluate the performance of group testing by conducting simulations. 
\end{abstract}


\section{Introduction}


Group testing is utilized in the case when we want to find a few defectives among large amount of items. 
For example, group testing can be applied to PCR tests for finding defective specimens, water quality test for the presence of harmful substances contained in wastewater from multiple locations, etc. 

Testing $n$ items one by one requires $n$ tests, but the ratio of defective items is often small (0.0001 to 0.01). 
In such cases, a test can be performed on a mixed pool of multiple items, and if the result is negative, it can be determined in a single test that all items in the pool are negative. 
If the result is defective, one of the items in the pool is defective. 
By testing various combinations of pools, the marginal posterior probability that each item is defective can be computed from the test results of a much smaller number of pools than the total number of items. 
However, the false positive/negative (FP/FN) probability of each test must be taken into account when making a positive/negative decision. 
Research on group testing originates from the syphilis testing by Robert Dorfman \cite{dorfman1943}. 
Group testing are classified into two types: noiseless testing (no FN/FP error) and noisy testing. 
Also, they are classified into two types: adaptive testing, in which the next pool is determined based on the results of the previous test, and nonadaptive testing, in which all pools are tested at once. 
Especially in the case when a single test is time-consuming, a large number of pools can be tested at once to identify defective items (\cite{du99}, \cite{aldridge19}).  
Algorithms such as Belief Propagation (BP) and MCMC are used for this purpose (\cite{knill96}\cite{uehara09}\cite{sejdinovic10}). 

\begin{figure}[h]
    \begin{center}
    \includegraphics[width=0.45\textwidth]{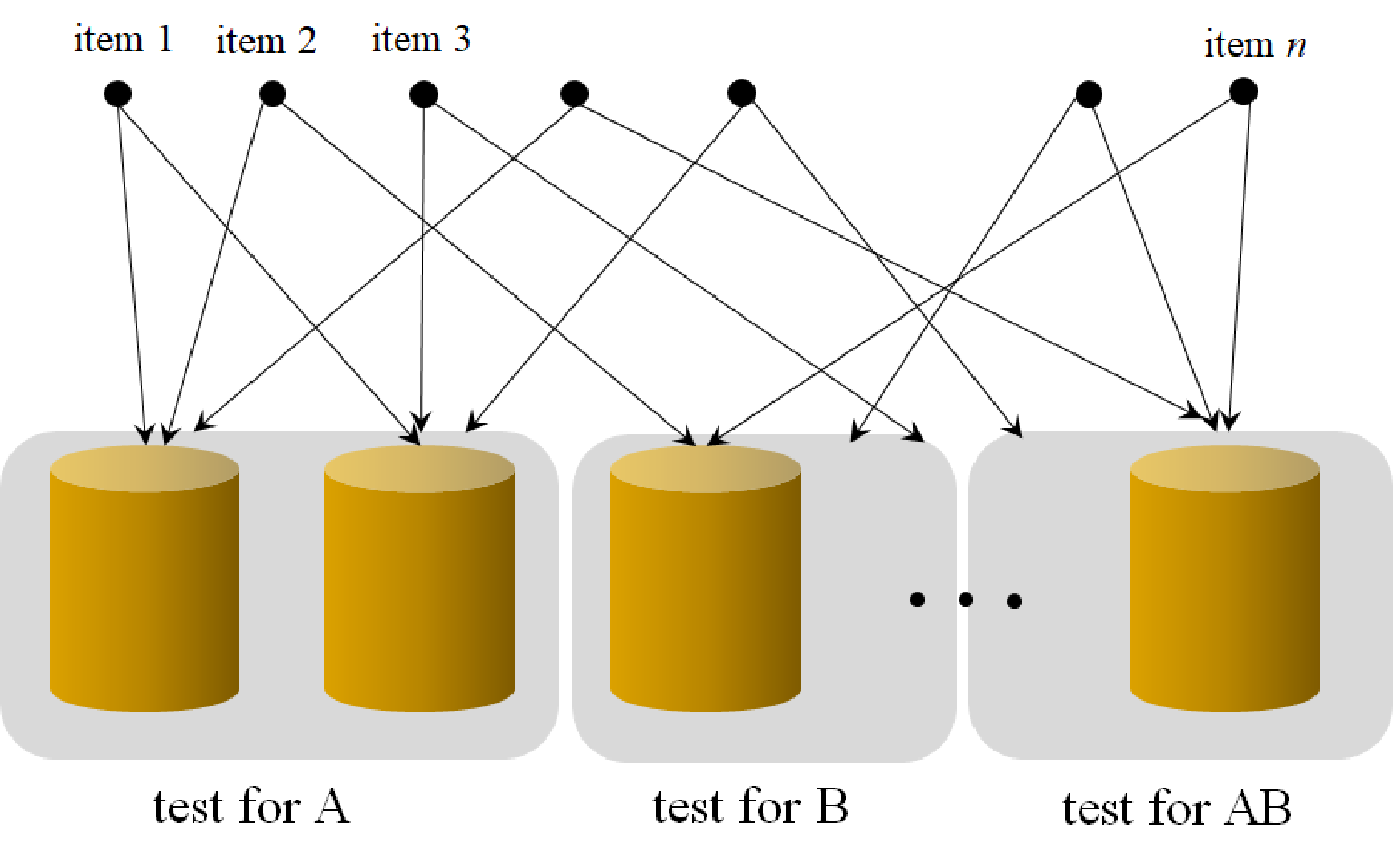}
    \vskip -5mm
    \caption{Group testing for two type defectives} 
    \label{fig: 200}
    \end{center}
\end{figure}

In this paper, we consider the case where two types of defective A and B exist. 
If group testing is conducted separately for each of A and B, the number of tests is twice as many as for one type's test. 
However, as shown in Fig.\ref{fig: 200}, it is expected that we can reduce the number of tests by mixing pools that perform tests for each of A and B, and pools that perform `test against AB' (i.e., a test where the test result is defective if either A or B is defective). 

For two types of defectives, firstly we construct a belief propagation algorithm to compute marginal posterior probabilities of defectives. 
Secondly we construct several kinds of collections of pools for tests for A, B, and AB by utilizing the finite affine geometry. 
And by utilizing our belief propagation algorithm we evaluate the performance of group testing by conducting simulations. 

\section{Group testing including two types of defectives}
In the case of two types of defectives, we construct three kinds of pools: (i)test for A, (ii)test for B, and (iii)test which reacts either A or B.

A set of various pools is called a pooling design. 
The combinatorial structure of a pooling design determines the efficiency of group testing. 

Let $C=\{c_1, \ldots, c_n\}$ be a set of items. 
Let $X_j^A=1$ if item $c_j$ is defective for type A and $X_j^A=0$ otherwise. 
Similarly $X_j^B$ is defined. 
Also, $X_j^{AB}=X_j^A \vee X_j^B$. 
A subset $G=\{c_{j_1}, \ldots, c_{j_k}\}$ of $C$ is called a pool. 
We simply write $1, \ldots, n$ to identify the elements of $C$ and their subscripts. And sometimes we write as $C=\{1, \ldots, n\}$, or, $G=\{j_1, \ldots, j_k\}$. 
Let $\cG^A$, $\cG^B$, and $\cG^{AB}$ denote the sets of pools that test for A, B, and AB, respectively, and
let $Z_i^A=\bigvee_{j \in G_i} X_j^A$ for the pool $G_i\in {\cG}^A$ that tests against A.
Let $\vect{Z}^A=(Z_i^A | G_i\in \cG^A)$. Define $\vect{Z}^B$ and $\vect{Z}^{AB}$ in the same way.
The observation $S_i^A$  for a pool of $G_i^A$ takes binary values 0, or 1, and let $\vect{S}^A=(S_i^A | i \in C)$.
The same applies to $\vect{S}^B, \vect{S}^{AB}$.
Sensitivity and specificity are defined by 
\begin{align*}
    p(1|1)&=\Pr(S_i^A=1 | Z_i^A=1)=\Pr(S_i^B=1 | Z_i^B=1)\\
    &=\Pr(S_i^{AB}=1 | Z_i^{AB}=1),\\
    p(0|0)&=\Pr(S_i^A=1 | Z_i^A=1)=\Pr(S_i^B=0 | Z_i^B=0)\\
    &=\Pr(S_i^{AB}=0 | Z_i^{AB}=0).
\end{align*}
That is, the probabilities $p(1|0)$ and $p(0|1)$ of FP and FN are assumed to be constant regardless of A, B and AB.

Let $p_A, p_B$ be the defective rates for type A and B. 
The marginal posterior probability that each item $c_j$ is positive/negative for A, B under the observations $\vect{s}^A$, $\vect{s}^B$, and $\vect{s}^{AB}$ is given by
\begin{equation}
    \Pr(X_j^A=a, X_j^B=b | \vect{s}^A, \vect{s}^B, \vect{s}^{AB})
    \label{eq: 73}
\end{equation}
for each case of $a,b =0, 1$.

Let $\vect{X}^A=(X_j^A | j \in C)$, $\vect{X}^B=(X_j^B | j \in C)$, and
let $\vect{X}_{-c}^A=(X_{c'}^A | c' \in C\setminus \{c\})$ be the vector of test results for items except for $c$.
Also, the event that $X_c^A=a$ for item $c$ and that $\vect{X}_{-c}^A=\vect{x}_{-c}^A$ for the rest are written by 
$\vect{X}^A=(\vect{x}_{-c}^A, a)$.
The similar notations are used for B. 
The marginal posterior probability can be written as follows: 
\begin{align*}
    \Pr(&X_j^A=a, X_j^B=b | \vect{s}^A, \vect{s}^B, \vect{s}^{AB})\\
    =&
    K \sum_{\vect{x}_{-j}^A, \vect{x}_{-j}^B} \Pr(\vect{X}^A=(\vect{x}_{-c}^A, a), \vect{X}^B=(\vect{x}_{-c}^B, b))\\
    &\times
    \Pr(\vect{S}=\vect{s} | \vect{X}^A=(\vect{x}_{-c}^A, a), \vect{X}^B=(\vect{x}_{-c}^B, b)),
\end{align*}
where $\vect{S}=(\vect{S}^A,\vect{S}^A,\vect{S}^{AB})$, and $K=\Pr(\vect{S}=\vect{s})^{-1}$.

This calculation involves  $2^{n-2}$ sums, and when $n$ is large, the computational complexity is $O(2^n)$.

\section{Belief propagation algorithm}

Pooling designs can be represented by three bipartite graphs $(C, \cG^A, E^A), (C, \cG^B, E^B), (C, \cG^{AB}, E^{AB})$ whose vertices are connected if a pool $G$ contains a item $c$ as shown in Fig. \ref{fig: 200}, where $E^A, E^B, E^{AB}$ are the sets of
edges of these bipartite graphs. And let $E=E^A\cup E^B\cup E^{AB}$.

In order to calculate the marginal posterior probability, we develop the following algorithm based on belief propagation for screening two types of defectives.

\begin{enumerate}
    \item[Step 1] (Initialization of $Q$): For each edge $(c, G) \in E$, let
        \[
        \bar{Q}_{cG}^{(0)}(x, y) := \Pr(X_c^A=x)\Pr(X_c^B=y) \quad (x, y) \in \{0, 1\}^2.
        \]
        Let $t:=1$ and let $\varepsilon > 0$.
    \item[Step 2] (Computation of $R$): When $G\in \cG^A$, for edge $(c, G)\in E^A$, and  for $y=0, 1$, let
        \begin{align*}
            R_{Gc}^{(t)}&(0, y):=
            p(s_G | 1)\\+&(p(s_G | 0) - p(s_G | 1)) \\
            &\times \prod_{c' \in G\setminus \{c\}} 
            \left(\bar{Q}_{c'G}^{(t-1)}(0, 0)+\bar{Q}_{c'G}^{(t-1)}(0, 1)\right)\\
            R_{Gc}^{(t)}&(1, y):=p(s_G | 1).
        \end{align*}
        When $G\in \cG^B$, for edge $(c, G)\in E^B$, and for $x=0, 1$, 
        \begin{align*}
            R_{Gc}^{(t)}&(x, 0):=
            p(s_G | 1)\\+&(p(s_G | 0) - p(s_G | 1)) \\
            &\times \prod_{c' \in G\setminus \{c\}} 
            \left(\bar{Q}_{c'G}^{(t-1)}(0, 0)+\bar{Q}_{c'G}^{(t-1)}(1, 0)\right)\\
            R_{Gc}^{(t)}&(x, 1):=p(s_G | 1).
        \end{align*}
        When $G\in \cG^{AB}$, for edge $(c, G)\in E^{AB}$, let
        \begin{align*}
            R_{Gc}^{(t)}&(0, 0):=
            p(s_G | 1)\\
            +&(p(s_G | 0) - p(s_G | 1)) 
            \prod_{c' \in G\setminus \{c\}} \bar{Q}_{c'G}^{(t-1)}(0, 0),\\
            R_{Gc}^{(t)}&(x, y):=p(s_G | 1), \quad (x, y) \ne (0, 0).
        \end{align*}
    \item[Step 3] (Computation of $Q$): For each edge $(c, G)\in E$, let
        \begin{align*}
        Q&_{cG}^{(t)}(x, y) \\&:= \Pr(X_c^A=x)\Pr(X_c^B=y) \prod_{G' \in (c)\setminus \{G\}} \bar{R}_{G'c}^{(t)}(x, y).
        \end{align*}
        Let $K_{cG}^{(t)}:=\sum_{x, y} Q_{cG}{(t)}(x, y)$, and normalize $\bar{Q}_{cG}{(t)}(x, y):=Q_{cG}{(t)}(x, y)/K_{cG}^{(t)}$.
    \item[Step4] (Repeat for each $t$) If \[\max_{c\in C}|\bar{Q}_{cG}^{(t-1)}(x, y) -\bar{Q}_{cG}^{(t)}(x, y) | < \varepsilon, \]
        then go to Step 5, otherwise, let $t:=t+1$ and go to Step 2
    \item[Step 5] (computation of marginal posterior probability): 
        For each $c \in C$, let
        \begin{align*}
        Q_c&(x, y)\\:=&\Pr(X_c^A=x, X_c^B=y)\\& \times \prod_{G\in (c)} \bar{R}_{Gc}^{(t)}(x, y), \quad (x, y) \in \{0, 1\}^2.
        \end{align*}
        Finally, let $K_{c}:=\sum_{x, y} Q_{c}{(t)}(x, y)$ and normalize $\bar{Q}_{c}(x, y):=Q_{c}(x, y)/K_{c}$.
\end{enumerate}

In this study, we use the above BP algorithm.

\section{Pooling Design}

In group testing, the `goodness' of the pooling design also affects the screening efficiency in addition to the effective algorithm for screening. 
Combinatorial properties such as $\bar{d}$-separable and $d$-disjunct have been studied in the case of noiseless testing with a single kind of defective.
(see, for example, Du et al. \cite{du99})

When a single type defective is taking into account, for a set of pools $\cG=\{G_1, \ldots, G_m\}$, a pair $(C, \cG)$ is called a pooling design. 
A pooling design is represented by an  matrix $M=(m_{ij})$, where
    \[
        m_{ij}= \begin{cases} 1 & \text{if item $c_j$ is included in pool $G_i$} \\ 0 & \text{otherwise}\end{cases}
    \]
Each row of $M$ corresponds to a pool and each column corresponds to an item. 
For an item $c_j$, $T_j=\{i |m_{ij}=1 \}$ is called the support of $c_j$. 
Let $\cT=\{T_1, \ldots, T_n\}$ be the set of supports. 
Given a positive integer $d$,  for any $0 \le d_i \le d$ ($i=1, 2$) and for any $T_1, \ldots, T_{d_1} \in \cT$ and $T_1^\prime, \ldots T_{d_2}^\prime \in \cT$, if
    \[
        T_1 \cup \cdots \cup T_{d_1} \ne T_1^\prime \cup \cdots\cup T_{d_2}^\prime
    \]
holds, then the pooling design is said to be $\bar{d}$-separable. 
In the case when we consider a single type defective and no FP/FN exist, that is, noiseless testing, if the pooling design is $\bar{d}$-separable, it is known that defective items less than $d$ can be accurately identified (see, for example, \cite{du99}). 

Also, given a positive integer $d$, for any distinct $d$ supports $T_1, \ldots, T_d \in \cT$ and $T_0\in \cT$ such that $T_0\ne T_i (i=1, \ldots ,d )$, if
    \[
        T_0 \not\subset T_1 \cup \cdots \cup T_d,
    \]
then the pooling design$(C, \cG)$ is said to be $d$-disjunct. 
It is known that if a pooling design is  $d$-disjunct, then it is $\bar{d}$-separable. 

For noisy testing with FP/FN, a pooling design with large $\bar{d}$-separability is expected to identify or screen more defective items. 

A similar combinatorial structure is defined in the case of two types A and B. 
Let $M_A, M_B, M_{AB}$ be  matrices of the pooling designs corresponding to each of the pools $\cG^A, \cG^B, \cG^{AB}$ that test for A, B, and AB, respectively. 
Let
\[
\overline{M}_A=\begin{pmatrix} M_A \\M_{AB}\end{pmatrix}, \quad \overline{M}_B=\begin{pmatrix} M_B \\M_{AB}\end{pmatrix}.
\]
Lu et al. \cite{lu2024} showed that if both $M_A$ and $M_B$ are $(d-1)$-disjunct and both $\overline{M}_A$ and $\overline{M}_B$ are $\bar{d}$-separable then all defective items can be correctly identified when the total number of defective items of A and B is less than $d$. 
This property is called $(2, \bar{d})$-separable. 

In a combinatorial sense, large separability is desired for indentifying defectives. 
However, if the probability of non-separable structures in a pooling design is less, it may have more ability of identifying defectives than its designed separability.
Under the usage of BP algorithm, we wish to investigate the relationship between the combinatorial structure of the pooling design and the identifiability or discriminability of defective items when two types of defective are included and FP/FN errors are present. 

\section{Construction of pooling design}
\label{sec:design}

It is known that BP algorithms converge to the exact marginal posterior probability if there are no cycles in the bipartite graphs of a pooling design. 
However, in the case of pooling designs of group testing, we can not avoid cycles in the bipartite graphs in order to construct efficient designs. 
If there are short cycles in the bipartite graphs of a pooling design, a BP algorithm do not converges to the accurate values, which have some errors, and often it does not even converge. 
Hence, a pooling design not having short cycles in the bipartite graphs are desired. 
The shortest cycle is length four in a bipartite graph. 

We want to avoid cycles of length four in bipartite graphs $(C, \cG^A, E^A), (C, \cG^B, E^B), (C, \cG^{AB}, E^{AB})$. 
In other words, it is desired that for any two rows (pools) in each $M_A$, $M_B$, $M_{AB}$, there is at most one column (item) which have 1's in common. 
This property is called the unique collinearity condition (see Uehara et al. \cite{uehara09}). 

In adittion to the property of separability or disjunctness, it is required that a pooling design should satisfy the unique collinearity
condition. A combinatorial design called `packing design' has this property. The following construction of a pooling design
satisfies the unique collinearity condition.

For a prime or a prime power $q$, let AG$(3, q)$ be the 3-dimensional affine geometry  over the finite field $\F_q=\{f_0, \ldots, f_{q-1}\}$.
Each point of AG$(3, q)$ is represented as $(y_0, y_1, y_2)$ ($y_i \in \F_q$) and AG$(3, q)$ consists of $q^3$ points. 
Let $P_i$ be a plane satisfying $y_0=f_i$ ($f_i \in \F_q$), each plane consists of $q^2$ points and $P_i$'s ($f_i \in \F_q$) form a parallel $q$ planes. 
Let $\cL$ be the set of lines that intersect each $P_i$ at exactly one point, and $|\cL|=q^4=n$. 
Let $\cG$  be the set of points corresponding to pools and let $C$ be the set of  lines corresponding to items. 
Then the incidence (or adjacency) matrix $M$ of the pools and items is determined. 

Partitioning $q^3$ points into $q$ planes $P_i$ ($f_i \in \F_q$), $q^2\times n$ incidence submatrices $M_i$ of points on $P_i$ and lines are obtained. 
Then $M$ consists of $M_i$'s vertically aligned in 

\[
  M=\begin{pmatrix} M_0 \\ \vdots\\M_{q-1}\end{pmatrix}.
\]

There are $q^2$ 1's in each row of $M_i$ and exactly one 1 in each column. %
Let $K \subset \F_q$ and $|K|=k$. 
By piling up $k (\le q)$ $M_i$'s ($i \in K$), we can make a $kq^2\times n$ incidence matrix $M_K$. 
Each column of $M_K$ has $k$ ones, thus it is $(k-1)$-disjunct. 

Assume that the probabilities of occurrence of the two type defectives A and B are equal, that is,  $p_A=p_B$. 
Using $M_i$'s in the above 3-dimensional affine geometry, we construct  incidence (or adjacency) matrices $M_A, M_B, M_{AB}$ as follows. 
For $K \subset \F_q$, let $M_A=M_B$ be a $kq^2\times n$ matrix with $M_i$'s ($ i \in K$) vertically aligned and $M_{AB}$ be a $(q-k)q^2\times n$ matrix with $M_i$'s ($ i \in K^c$) vertically aligned. 

In our simulation, the $M_A, M_B, M_{AB}$ constructed above are used. 

\section{Simulation Results}
\label{sec:result}

In our simulations FP and FN are fixed as $p(1 | 0)=0.01$ and $p(0 | 1)= 0.03$, respectively.
And assume that $p_A=p_B=0.002$. 
We use pooling designs generated by AG$(3, 7)$ by setting $q=7$. 
In a simulation, each item is tested $k$ times, $k$ times, and $7-k$ times, respectively for A, B, and AB. 
The simulations are executed for $k=1, 2, \ldots, 6$ by utilizing corresponding  matrices $M_A, M_B, M_{AB}$. 
We use distinct $M_i$'s between $M_A$ and $M_{AB}$, also, between $M_B$ and $M_{AB}$. 
However, they may not distinct between $M_A$ and $M_B$ since from the separability point of view it is allowed that $M_A=M_B$. 

Each simulation is repeated  1000 times. 
Table \ref{tbl: 97} shows the worst rank that all true defectives are included with probabilities 95\% and 99\%  when the marginal posterior probabilities of each item being defective are sorted in descending order. 
In the table, note that the first row shows the number of defectives for each of A and B. 

Among the simulations, design (3) reveals the most effective results. 
Even if the number of defectives are 10 for each of A and B, the screening or identification powers are still high. 
On the other hand designs (1), (5), (6) has lower identifiability. 
In the case of (1), the number of individual tests for each item is one in $M_A$ and $M_B$, respectively. 
It means that more replication is required for the separability property. 
Designs (5) and (6) has low screening power for probability 99\% in the case of 8 or 10 defectives. 
In these cases low screening power may due to the short cycles between $M_A$ and $M_B$. 

From the separability point of view, it is allowed that $M_A$ and $M_B$ are identical, but when a BP algorithm is adopted it is required that $M_A$ and $M_B$ are not be identical.
\begin{table*}[tp]
    \centering
    \caption{Ranking that includes all defectives for A and B}
    \label{tbl: 97}
    \begin{tabular}{c|rr|rr|rr|rr|rr|rr}
        \hline
        \# of defectives & \multicolumn{2}{c|}{2} & \multicolumn{2}{c|}{4} & \multicolumn{2}{c|}{6} & \multicolumn{2}{c|}{8} & \multicolumn{2}{c|}{10} & \multicolumn{2}{c}{12} \\ \hline
         \diagbox{design}{probability of \\screening} & 99\% & 95\% & 99\% & 95\% & 99\% & 95\% & 99\% & 95\% & 99\% & 95\% & 99\% & 95\% \\ \hline
        (1) $m_A=m_B=49$,  &   A:3 &  A:2 & A:6 &  A:5 &  A:9 &  A:8 &  A:12 &  A:11 &  A:21 & A:15 & A:453 & A:112 \\ 
            $m_{AB}=294$   &   B:3 &  B:3 & B:6 &  B:5 &  B:9 &  B:8 &  B:14 &  B:11 &  B:22 & B:15 & B:307 & B:101 \\ \hline
        (2) $m_A=m_B=98$,  &   A:2 &  A:2 & A:4 &  A:4 &  A:6 &  A:6 &   A:9 &   A:8 &  A:12 & A:11 & A:100 &  A:24 \\ 
            $m_{AB}=245$   &   B:2 &  B:2 & B:4 &  B:4 &  B:6 &  B:6 &   B:9 &   B:8 &  B:12 & B:11 & B:104 &  B:18 \\ \hline
        (3) $m_A=m_B=147$, &   A:2 &  A:2 & A:4 &  A:4 &  A:6 &  A:6 &   A:9 &   A:8 &  A:14 & A:10 &  A:21 &  A:13 \\ 
            $m_{AB}=196$   &   B:2 &  B:2 & B:4 &  B:4 &  B:6 &  B:6 &   B:9 &   B:8 &  B:13 & B:10 &  B:18 &  B:13 \\ \hline
        (4) $m_A=m_B=196$, &   A:2 &  A:2 & A:4 &  A:4 &  A:6 &  A:6 &  A:12 &   A:8 &  A:19 & A:10 &  A:32 &  A:14 \\ 
            $m_{AB}=147$   &   B:2 &  B:2 & B:4 &  B:4 &  B:7 &  B:6 &  B:12 &   B:8 &  B:19 & B:10 &  B:29 &  B:13 \\ \hline
        (5) $m_A=m_B=245$, &   A:2 &  A:2 & A:4 &  A:4 &  A:7 &  A:6 &  A:16 &   A:8 &  A:52 & A:16 & A:107 &  A:15 \\ 
            $m_{AB}=98$    &   B:2 &  B:2 & B:4 &  B:4 & B:11 &  B:6 &  B:27 &   B:8 &  B:88 & B:10 & B:132 &  B:13 \\ \hline
        (6) $m_A=m_B=294$, &   A:2 &  A:2 & A:4 &  A:4 & A:16 &  A:6 &  A:88 &   A:8 & A:191 & A:10 & A:523 &  A:17 \\ 
            $m_{AB}=49$    &   B:2 &  B:2 & B:4 &  B:4 & B:23 &  B:6 &  B:66 &   B:8 & B:403 & B:10 & B:557 &  B:15 \\ \hline
    \end{tabular}
\end{table*}

\section{Conclusion}
This paper address to the case where two types of defective A and B exist in group testing.
In order to screening two types of defectives, we developed a belief propagation algorithm to compute marginal posterior probability of defectives. 
By utilizing the finite affine geometry, we construct several kinds of collections of pools for test A and B.
And simulation is conducted to evaluate performance of group testing by using our belief propagation algorithm.

Through simulation experiments on the adopted pooling design, we suggest the follows; 
proposed BP algorithm shows high screening performance for two types of defectives when the number of each defectives is about 8. 
The identification power of the proposed BP algorithm decreases when there is a cycle structure on the adopted pooling design. 
Future work is needed to improve the pooling design and the BP algorithm.


\bibliographystyle{IEEEtran}
\bibliography{IEEEabrv,reference}

\end{document}